\documentclass{webofc}
\usepackage[varg]{txfonts}   
\usepackage{caption}
\usepackage{subcaption}
\usepackage{amsmath}
\usepackage{amssymb}
\usepackage{color}
\usepackage{graphicx,url}
\usepackage[english]{babel}
\usepackage[utf8]{inputenc}

\newcommand{\im}{\mbox{Im}\,}
\newcommand{\re}{\mbox{Re}\,}
\newcommand{\Od}{{\mathcal O}}
\newcommand{\T}{\mathcal{T}}

\begin{document}
\title{Pion scattering in finite volume within the Inverse Amplitude Method: Extension to moving frames}

\author{\firstname{Julián A.} \lastname{Sánchez}\inst{1}\fnsep\thanks{\email{Julian.Sanchez@ific.uv.es}} \and
        \firstname{Raquel} \lastname{Molina}\inst{1}\fnsep\thanks{\email{Raquel.Molina@ific.uv.es}} \and
        \firstname{Angel} \lastname{Gómez Nicola}\inst{2}\fnsep\thanks{\email{gomez@fis.ucm.es}}
}

\institute{Departamento de F\'isica Te\'orica and IFIC, Centro Mixto Universidad de Valencia-CSIC, Parc Cient\'ific UV, C/ Catedr\'atico Jos\'e Beltr\'an, 2, 46980 Paterna, Spain \and  Universidad Complutense de Madrid, Facultad de Ciencias F\'isicas, Departamento de F\'isica Te\'orica and
IPARCOS, Plaza de las Ciencias 1, 28040 Madrid, Spain}

\abstract{
We study the leading exponentially suppressed finite-volume corrections to $\pi\pi$ scattering near threshold in moving frames. Using the one-loop Inverse Amplitude Method (IAM), we compute the $I=0$ and $I=2$ amplitudes in a box with total momentum $\vec P=2\pi\vec d/L$, including the left-hand cut and the $t$-, $u$-channel and tadpole contributions neglected in Lüscher's quantization condition. We derive the resulting correction to the near-threshold energy levels and measure its size relative to the Lüscher term. We find that it becomes comparable to the latter for $m_\pi L\lesssim 1.5$--$1.7$ at rest, and that it is larger in moving frames, reaching $m_\pi L\lesssim 2.8$ for $\vec d=[002]$. The corrections become negligible only for $m_\pi L\gtrsim 5$.
}
\maketitle
\section{Introduction}
\label{intro}

Lattice QCD allows for a first-principles study of hadron scattering and of resonances such as the $\sigma$ and the $\rho$~\cite{RefA, RefB, Refoller, RefC}. 
Resonance properties are extracted from the finite-volume spectrum through Lüscher's quantization condition~\cite{RefA}, which can be extended to moving frames with total momentum $\vec P=2\pi\vec d/L$ to access more energy levels~\cite{RefE,RefF}. This formalism, however, neglects finite-volume effects that are exponentially suppressed in $m_\pi L$, such as those from the left-hand cut (LHC) and from $t$-, $u$-channel and tadpole loops~\cite{RefD,RefE}. They vanish for large boxes but may become relevant for small $m_\pi L$, and they are expected to depend on the boost.

To quantify them, we use Chiral Perturbation Theory (ChPT) unitarized with the one-loop Inverse Amplitude Method (IAM)~\cite{Refoller,RefC}. The IAM was formulated in a finite volume in Ref.~\cite{RefF}; here we extend that analysis to moving frames and study the near-threshold $\pi\pi$ amplitude in the $I=0$ and $I=2$ channels for several boosts $\vec d$.

We first review the ChPT and IAM amplitudes in infinite volume (Sec.~\ref{sec-1}), then present the finite-volume amplitude and the modified quantization condition (Sec.~\ref{sec-2}), and finally discuss the size of the corrections (Sec.~\ref{sec-3}). Conclusions are given in Sec.~\ref{sec-4}.

\section{The ChPT and IAM amplitudes at infinite volume}
\label{sec-1}

The elastic pion scattering amplitude $\pi^a \pi^b \rightarrow \pi^c \pi^d$, denoted $\T_{abcd}$, depends on the Mandelstam variables $s,t,u$ built from the external momenta, with total four-momentum $P$. Isospin invariance and crossing symmetry imply that only one function $A(s,t,u)$ is needed to characterize any pion scattering amplitude, namely
\begin{align}
\T_{abcd}(s,t,u) &= A(s,t,u)\,\delta_{ab}\delta_{cd} + A(t,s,u)\,\delta_{ac}\delta_{bd} + A(u,t,s)\,\delta_{ad}\delta_{bc}.
\label{crossinginf}
\end{align}

The IAM~\cite{Refoller,RefC} approximates $\re(1/t)$, with $t$ the partial-wave projection of the amplitude, by its ChPT expansion up to $\Od(p^4)$, i.e.\ $\re(1/t)=1/t_2 - \re\, t_4/t_2^2+\Od(p^6)$. Recall that $t_2$ is real for real $s$ and that perturbative unitarity implies $\im\, t_4=\sigma t_2^2$ above threshold, where $\sigma=\sqrt{1-4m_\pi^2/s}$ is the phase-space factor. The unitarized amplitude for partial waves then reads
\begin{equation}
t_{\rm mIAM}(s)=\frac{t_2^2 (s)}{t_2(s)-t_4(s) + A_{z}(s)},
\label{IAMsingle}
\end{equation}
where $A_z(s)$ is the correction that accounts for the Adler zero. 

\section{The ChPT scattering amplitude at finite volume}
\label{sec-2}

We consider a cubic box of side $L$ in a moving frame with total momentum $\vec P=2\pi\vec d/L$. The scattering amplitude is defined as the amputated four-point Green function in finite volume, with the external legs given by the dispersion relation of the renormalized propagator. In finite volume all spatial momenta, internal (in loops) or external, are discretized, i.e.\ a generic three-momentum is $\vec q=(2\pi/L)\,\vec n$ with $\vec n\in\mathcal{P}_{\vec d}$,
\begin{equation}
    \mathcal{P}_{\vec{d}} = \left\{ \vec{n}\;\Big|\; \vec{k} = \frac{2 \pi}{L}\, \boldsymbol{\gamma}^{-1}\Big(\vec{n} - \frac{1}{2}\, \vec{d}\, \epsilon \Big) \right\}, \qquad \epsilon = 1 + \frac{m_{1}^{2} - m_{2}^{2}}{s},
\end{equation}
where $\boldsymbol{\gamma}$ is the Lorentz-boost matrix and $\gamma=E/\sqrt{s}$ its Lorentz factor, with $E$ the total energy in the moving frame. 
For two pions of equal mass, $\epsilon=1$. Loop integrals become discrete sums over the spatial components of the internal momenta,
\begin{equation}
(-i)\int \frac{d^4 q}{(2\pi)^4}\rightarrow (-i)\frac{1}{L^3}\sum_{\vec{n}}\int \frac{d q_0}{2\pi}.
\label{sum}
\end{equation}

The finite-volume generalization of the crossing relation and of the isospin projections is
\begin{align}
    \T^0(S,T,U;L) &= 3A(S,T,U;L) + A(T,S,U;L) + A(U,T,S;L), \nonumber\\
    \T^1(S,T,U;L) &= A(T,S,U;L) - A(U,T,S;L), \nonumber\\
    \T^2(S,T,U;L) &= A(T,S,U;L) + A(U,T,S;L),
\label{isospin}
\end{align}
where $S^{\mu}=P^{\mu}$, $T^{\mu}=(p-p')^{\mu}$ and $U^{\mu}=(p+p'-P)^{\mu}$ are generalized Mandelstam variables, $p$ and $p'$ are the incoming and outgoing momenta, and $A$ is the finite-volume amplitude for $\pi^+\pi^-\to\pi^0\pi^0$. In finite volume the amplitude depends on the individual momenta and not only on $s,t,u$; this dependence is kept implicit here.

We write the amplitude for each isospin channel as
\begin{equation}
\T^I(S,T,U;L)=\T^I(s,t,u;L\rightarrow\infty) + \Delta \T^I(S,T,U;L).
\end{equation}
All UV divergences are contained in the $L\rightarrow\infty$ term, so the amplitude is rendered finite by the standard infinite-volume renormalization in dimensional regularization.

At one loop, the finite-volume contributions can be written as
\begin{equation}
    \T_{4} = \T_{4}^{\infty} + \sum_{i = H, u,t,2u, 2t} f_{i}\, \Delta J_{i},
\end{equation}
where $\Delta J_i$ are the exponentially suppressed finite-volume parts of the loop functions and $f_i$ are their coefficients in the amplitude, with $i = \{H, u, t, 2u, 2t\}$ ($H$ is tadpole contribution, and the other one are associated to the $t-$ and $u-$channel loops, see in \cite{RefF} for more details). The $s$- and $2s$-channel loops are absorbed in the exponentially suppressed part $\Delta J^{\rm exp}(s)$ and in the universal Lüscher function $\mathcal{S}_d$ defined below. Absorbing $\Delta J^{\rm exp}(s)$ in $\delta\T_4$ (see Ref.~\cite{RefF} for details), $\delta\T_4$ contains the $t$- and $u$-channel and tadpole finite-volume corrections together with the exponentially suppressed part of the $s$-channel. Near threshold, the quantization condition can then be reduced to
\begin{equation}
    p(s) \cot{\delta} - 16 \pi \sqrt{s}\, \frac{\delta \T_{4}(s;L)}{\T_{2}^{2}} = \frac{1}{\gamma\pi L}\, \mathcal{S}_{d}\left(\frac{p(s)^{2} L^{2}}{4 \pi^2}\right),
\label{eq:luscher-like}
\end{equation}
where $\delta(s)$ is the infinite-volume phase shift, related to the full infinite-volume amplitude as usual,
\begin{equation}
    T^{\rm IAM}(s) = \frac{\sqrt{s}}{2}\Big(p(s)\cot{\delta(s)} - i\, p(s) \Big)^{-1},
\end{equation}
and $\mathcal{S}_{d}$ is the Lüscher function,
\begin{equation}
    \mathcal{S}_{d}\left(\frac{p^2(s) L^2}{4\pi^2}\right)=4\pi^2 L\left[  \frac{1}{L^{3}} \sum_{\vec{q} \in \mathcal{P}_{\vec d}} - \int \frac{d^{3}\vec{q}}{(2 \pi)^{3}}\right]
    \frac{1}{\vert\vec{q}\vert^2-p^2(s)}.
\label{SLuscher}
\end{equation}

Expanding around threshold, $E\simeq 2m_\pi$ and $p\cot\delta\simeq 1/a$, with $a$ the continuum scattering length (which includes the LEC contributions), Eq.~\eqref{eq:luscher-like} gives the finite-volume energy shift $\Delta E=E_0-2m_\pi$ of the lowest level $E_0$. The additional near-threshold contribution from $\delta\T_4$ is measured by
\begin{equation}
    \mathcal{P} = 32\pi m_{\pi}\,
    \left| a\, \frac{\delta \T_{4}}{\T_{2}^{2}} \right|_{s \to 4m_{\pi}^{2}}.
\label{Pdef}
\end{equation}
The Lüscher limit corresponds to $\mathcal{P}=0$, while $\mathcal{P}\simeq 1$ means that the $\delta\T_4$ corrections are of the same order as the Lüscher term.

\section{Results and discussion}
\label{sec-3}

To estimate the size of the different contributions to the amplitude, we consider the $I=0,2$ projections at threshold in the center-of-mass frame, i.e.\ for $E=2m_{\pi}$ and $\vec{p}=\vec{p}\,'=\vec{0}$. In that limit one finds closed analytic expressions for the finite-volume amplitude and the energy levels, obtained as in Ref.~\cite{RefF}, which allow a direct comparison of the $s$-, $t$-, $u$-channel and tadpole contributions. We use the physical pion mass and the LECs of Ref.~\cite{Hanhart:2008mx}.

\begin{figure}
     \centering
     \begin{subfigure}[b]{0.90\textwidth}
         \centering
         \includegraphics[width=\textwidth]{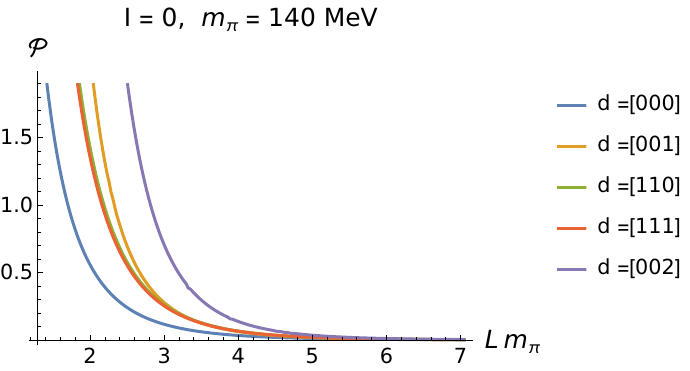}
         \caption{$\mathcal{P}$ for $I = 0$}
         \label{fig:EI0}
     \end{subfigure}
     \hfill
     \begin{subfigure}[b]{0.90\textwidth}
         \centering
         \includegraphics[width=\textwidth]{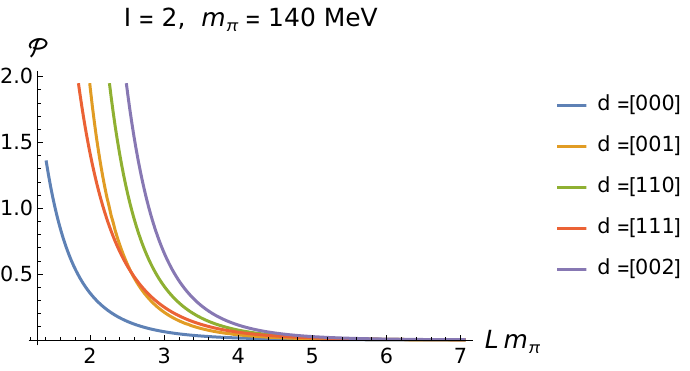}
         \caption{$\mathcal{P}$ for $I = 2$}
         \label{fig:EI2}
     \end{subfigure}
        \caption{Quantity $\mathcal{P}$ of Eq.~\eqref{Pdef}, which measures the finite-volume corrections to the Lüscher approach near threshold, as a function of $m_\pi L$ for different boosts $\vec d$ and for $I=0,2$. We use the LECs of Ref.~\cite{Hanhart:2008mx}; the uncertainty bands correspond to the variation of the renormalized couplings $l_{1-4}^r$.}
        \label{fig:1}
\end{figure}

The relevant quantity is $\mathcal{P}$ of Eq.~\eqref{Pdef}: $\mathcal{P}=0$ is the Lüscher limit, while $\mathcal{P}\simeq 1$ means that the exponential corrections are of the same order as the Lüscher term. Fig.~\ref{fig:1} shows $\mathcal{P}$ as a function of $m_\pi L$ for each boost, and Table~\ref{tab:P1} gives the value of $m_\pi L$ below which $\mathcal{P}\gtrsim 1$. At rest, the corrections, although formally exponentially suppressed, are already comparable to the Lüscher term for $m_\pi L\lesssim 1.5$--$1.7$. In moving frames they reach this size at larger volumes, $m_\pi L\lesssim 2.15$--$2.54$ for $\vec d=[001],[110],[111]$ and $m_\pi L\lesssim 2.8$ for $\vec d=[002]$ ($\vec d^{\,2}=4$), so boosted frames are more sensitive to them. For $m_\pi L\gtrsim 5$ we find $\mathcal{P}\to 0$ in all cases. 

\begin{table}
\centering
\begin{tabular}{lcc}
\hline
$\vec d$ & $I=0$ & $I=2$ \\
\hline
$[000]$ & $1.70$ & $1.52$ \\
$[001]$ & $2.33$ & $2.25$ \\
$[110]$ & $2.15$ & $2.54$ \\
$[111]$ & $2.15$ & $2.15$ \\
$[002]$ & $2.80$ & $2.78$ \\
\hline
\end{tabular}
\caption{Value of $m_\pi L$ below which $\mathcal{P}\gtrsim 1$, for each boost $\vec d$ and isospin channel (threshold evaluated in the center-of-mass frame).}
\label{tab:P1}
\end{table}

\section{Conclusions}
\label{sec-4}

We have extended the finite-volume IAM analysis of $\pi\pi$ scattering to moving frames, computing the leading exponential corrections to the amplitude near threshold in the $I=0$ and $I=2$ channels. These corrections modify Lüscher's quantization condition through the term $\delta\T_4$, whose relative size is measured by the quantity $\mathcal{P}$ of Eq.~\eqref{Pdef}.

We find that, although formally suppressed exponentially, the corrections become comparable to the Lüscher term ($\mathcal{P}\simeq 1$) for $m_\pi L\lesssim 1.5$--$1.7$ at rest. In moving frames the exponential contributions are more important than at rest, $\vec d=[002]$ ($\vec d^{\,2}=4$) being the most sensitive case, where $\mathcal{P}\simeq 1$ is reached for $m_\pi L\lesssim 2.8$. The exponential contributions become negligible only for $m_\pi L\gtrsim 5$.

As future work, we plan to compute the energy levels beyond threshold and the corresponding phase shifts, including the $\rho$ channel ($I=1$), and to study the effect of the pion mass on the extracted observables in the moving-frame setup.

%
%

\section*{Acknowledgments}
\label{sec:thanks}

We thank Michael Döring and Raúl A. Briceño for useful discussions.  This work is supported by the Ministerio de Ciencia e Innovación, research contracts PID2022-136510NB-C31 and PID2023-147458NB-C21,
funded by MICIU/AEI/10.13039/501100011033. R. M. also acknowledges support from the ESGENT program with Ref. ESGENT/018/2024 and the PROMETEU program with Ref. CIPROM/2023/59, of the Generalidad Valenciana (GVA), and also from 
the Spanish Ministerio de Economia y Competitividad and European Union (NextGenerationEU/PRTR) by the grant with Ref. CNS2022-136146. R. M. and J. A. S. thank to the Grisolia Program of the Consejería de Innovación, Universidades, ciencia y sociedad digital de la GVA, proyecto con referencia  CIGRIS/2021/089. This work is also partly supported by the Spanish Ministerio de Economia y Competitividad (MINECO) and
European FEDER funds under Contracts No. FIS2017-84038-C2-1-P B, PID2020-112777GB-I00, and by Generalitat Valenciana under contract PROMETEO/2020/023. This project has received funding from the European Union Horizon 2020 research
and innovation program under the program H2020-INFRAIA-2018-1, grant agreement No. 824093 of the STRONG-2020.


\begin{thebibliography}{}

\bibitem{RefA}
M.~L\"uscher, Commun. Math. Phys. \textbf{105}, 153 (1986).

\bibitem{RefB}
arXiv:1205.35082v1 [hep-lat] (2012).

\bibitem{Refoller}
J.~A.~Oller, E.~Oset and J.~R.~Pel\'aez, Phys. Rev. D \textbf{59}, 074001 (1999).

\bibitem{RefC}
A.~G\'omez Nicola and J.~R.~Pel\'aez, Phys. Rev. D \textbf{65}, 054009 (2002).

\bibitem{RefD}
P.~F.~Bedaque, I.~Sato and A.~Walker-Loud, Phys. Rev. D \textbf{73}, 074501 (2006).

\bibitem{RefE}
M.~D\"oring, H.-W.~Hammer, M.~Mai, J.-Y.~Pang, A.~Rusetsky and J.~Wu, Phys. Rev. D \textbf{97}, 114508 (2018).

\bibitem{RefF}
A.~G\'omez Nicola, R.~Molina and J.~A.~S\'anchez, JHEP \textbf{06}, 057 (2026), doi:10.1007/JHEP06(2026)057 [arXiv:2512.23462 [hep-lat]].

\bibitem{Hanhart:2008mx}
C.~Hanhart, J.~R.~Pel\'aez and G.~R\'ios, Phys. Rev. Lett. \textbf{100}, 152001 (2008).

\end{thebibliography}
\end{document}